\documentclass[12pt]{article}

\newcommand{\be}{\begin{equation}}
\newcommand{\ee}{\end{equation}}

\begin{document}
\title{{\bf PHASE TRANSITIONS FOR
\\ GAUGE THEORIES ON TORI
\\ FROM THE ADS/CFT CORRESPONDENCE}
\thanks{Alberta-Thy-02-07, hep-th/0205001}}
\author{
Don N. Page
\thanks{Internet address:
don@phys.ualberta.ca}
\\
CIAR Cosmology Program, Institute for Theoretical Physics\\
Department of Physics, University of Alberta\\
Edmonton, Alberta, Canada T6G 2J1
}
\date{(2002 April 30)}
\maketitle
\large

\begin{abstract}
\baselineskip 16 pt

	The vacuum of a large-$N$ gauge field on a $p$-torus
has a spatial stress tensor with tension along the direction
of smallest periodicity and equal pressures
(but $p$ times smaller in magnitude) along the other directions,
assuming an AdS/CFT correspondence and
a refined form of the Horowitz-Myers positive-energy conjecture.
For infinite $N$, the vacuum exhibits a phase transition when the lengths
of the two shortest periodicities cross.
A comparison is made with the Surya-Schleich-Witt
phase transition at finite temperature.
A zero-loop approximation is also given for large but finite $N$.

\end{abstract}
\normalsize
\baselineskip 16 pt
\newpage

	Horowitz and Myers
\cite{HM}
have noted that one can calculate the Casimir energy density
of a nonsupersymmetric Yang-Mills gauge theory
on $S^1\times R^p$,
in the limit that the number $N$ of gauge fields is made very large,
by using the AdS/CFT correspondence
\cite{mal,gkp,W}
and the supergravity energy of what they call the AdS soliton.
Here I note that if one takes the gauge theory to be defined on
$R^1\times T^p$, the product of a temporal $R^1$ with a spatial $p$-torus
(the product of $p$ $S^1$ circles),
then in the large-$N$ limit one gets vacuum phase transitions
(e.g., in the stress tensor)
when one varies the lengths of the $S^1$'s so that the values
of the two shortest lengths cross.
For a thermal state of the gauge theory at temperature $T$, which corresponds
to making the Euclidean time periodic with period $\beta = 1/T$,
so that the theory is defined on a Euclidean $(p+1)$-torus,
there is an additional phase transition,
found previously by Surya, Schleich, and Witt
\cite{SSW},
when $\beta$ crosses the length
of the shortest other period.

	The AdS soliton metric in $p+2$ spacetime dimensions is
\cite{HM}
\be
ds^2 ={r^2\over \ell^2}\left[\left (1 -{r_0^{p+1}\over r^{p+1}}\right)
 d\tau^2 + \sum_{i=1}^{p-1}(dx^i)^2 -dt^2  \right] +
 \left(1 -{r_0^{p+1}\over  r^{p+1}}\right)^{-1} {\ell^2\over r^2}dr^2,
\label{1}
\ee
where the radial variable $r$ ranges from $r_0$ to $\infty$,
and where to avoid a conical singularity at $r=r_0$,
one must make $\tau$ periodic with period $\beta_{\tau} = 4\pi\ell^2/ (p+1)r_0$.
(I have added the subscript $\tau$ to what Horowitz and Myers call simply
$\beta$ in order to distinguish that period, of the spatial coordinate $\tau$,
from my use of $\beta$ to denote the period of the Euclidean time
coordinate when I consider a thermal state.)

	This soliton, an Einstein metric with cosmological constant
$\Lambda=-p(p+1)/(2\ell^2)$ that is negative,
represents a solution for a supergravity
theory in $p+2$ dimensions in the classical limit
$\ell \gg\ell_{\mathrm Planck}$.
By the AdS/CFT correspondence, it should be dual to a suitable state
(e.g., the vacuum) of a large-$N$ gauge theory
defined on the conformal boundary of the AdS soliton metric, at $r=\infty$.

	Horowitz and Myers
\cite{HM}
considered the case in which $\tau$ was periodic but the other spatial
coordinates at constant $r$, the $p$ $x^i$'s, were not (except when
they normalized the energy, which is infinite for unbounded $x^i$'s).
Then with $t$ also unbounded, the dual gauge theory was defined on
$S^1\times R^p$ (with spatial sections $S^1\times R^{p-1}$).
However, I shall take the case in which each $x^i$ is periodic, with period
$L_i$ for $i=1,\cdots, p-1$.  For symmetry of notation, I shall also define
$x^p = \tau$ and $L_p = \beta_{\tau}$,
so each of the $p$ spatial coordinates for the gauge theory has period $L_i$,
but now with $i=1,\cdots, p$.  Thus the spatial part of the manifold
on which the gauge theory lives is the product of $p\,$ $S^1$'s, the $p$-torus
$T^p$.  I shall also take the case in which all of the fermionic fields
are antiperiodic around each of the $S^1$'s, so that in principle any of the
$S^1$'s could have length shrunk to zero at some locations
in the metric of the dual supergravity theory
(and hence representing a rotation by $2\pi$ at those locations,
reversing the sign of fermionic fields).

	Then in the case in which one is interested in the Lorentzian gauge
theory (so that the Lorentzian time $t$ has infinite range,
giving an $R^1$ factor), the total spacetime topology on which the gauge theory
lives is $R^1\times T^p$.  Up to an arbitrary (smooth, positive)
conformal factor, the metric of this spacetime is what is obtained from
the soliton metric (\ref{1}) by dropping the $dr^2$ part, multiplying
by a conformal factor $l^2/r^2$, and taking the limit $r \rightarrow \infty$:
\be
ds^2 = - dt^2 + \sum_{i=1}^{p-1}(dx^i)^2 + d\tau^2
     = - dt^2 + \sum_{i=1}^p(dx^i)^2.
\label{2}
\ee
This metric is of course flat, and the only nontrivial continuous parameters
are the $p$ lengths $L_i$ of the $S^1$ factors.
Since the gauge theory in this $p+1$ dimensional spacetime,
dual to the supergravity theory in $p+2$ dimensions,
is conformally invariant, only the $p-1$ ratios of the lengths are
physically relevant for conformally invariant properties of that theory.

	When the lengths $L_i$ are all multiplied by the same positive number
$c$, the conformally invariant gauge theory has the same physical form.
Since its energy $E$ has the dimension of inverse length,
it would be multiplied by $1/c$ under this scale transformation.
Thus the actual value of the energy of a CFT is not invariant
under conformal transformations.
However, when a representative of the conformal class of metrics for the CFT
is stationary, as is the metric (\ref{2}), and when any other external field
coupling to the CFT is also stationary (none in our example),
then in that representative metric the energy is well defined
and simply scales as $1/c$ if the lengths in the metric are scaled by $c$
under a constant conformal factor.
Therefore, if the energy is multiplied by a length scale taken from the metric,
the resulting product is invariant under the scaling.

	In our case we can use the spatial volume to define a length scale $L$.
If we follow Horowitz and Myers
\cite{HM}
to define $V_{p-1}$ to be the volume of their $p-1$ $x^i$'s,
i.e., $V_{p-1} = L_1L_2\cdots L_{p-1}$,
we can analogously define $V_p$ to be the volume of our $p$ $x^i$'s,
i.e.,
\be
V_p \equiv L^p = L_1L_2\cdots L_{p-1}L_p = V_{p-1}\beta_{\tau},
\label{3}
\ee
where the length scale $L$ is thus defined to be the geometric
mean of the $p$ spatial periodicities.
Then the scale-invariant quantity that reduces to the energy $E$
when the spatial volume is scaled to unity is
\be
\epsilon = E L \equiv E V_p^{1/p},
\label{4}
\ee
which I shall call the scale-invariant energy.

	Now by the AdS/CFT correspondence, one can equate
the energy of the CFT, for some choice of scale, with
the energy of the supergravity solution at the same choice of scale.
Using Eq. (3.16) of Horowitz and Myers
\cite{HM}
for the latter, one can readily calculate that the scale-invariant energy is
\be
\epsilon = -C_p \left({L\over\beta_{\tau}}\right)^{p+1}
     = -C_p \left({L\over L_p}\right)^{p+1},
\label{5}
\ee
\be
C_p \equiv \left({4\pi \over p+1}\right)^{p+1} {\ell^{\,p}\over 16 \pi G_{p+2}}
     = {1 \over 4(p+1)G_{p+2}}
         \left({8\pi^2 p\over (p+1)(-\Lambda)}\right)^{p\over 2}\, .
\label{6}
\ee

	This value comes from using the metric (\ref{1}), in which
it is the special coordinate $x^p = \tau$,
with coordinate periodicity $L_p = \beta_{\tau}$,
that has a proper length whose ratio with the proper length
of each other $p-1$ $x^i$
changes with $r$ and goes to zero at $r=r_0$.
In particular, the $p-1$ periodic $x^i$'s for $i=1,\cdots, p-1$
give circles whose proper lengths change in the same ratio as $r$
is reduced from $\infty$ to $r_0$, and whose proper lengths never go to zero,
but the proper length of the circle represented by $x^p$ changes at
a different rate with $r$ and goes to zero at the nut
\cite{GH}
at $r=r_0$, a regular center of polar coordinates for the $(r,x^p)$ two-surface.

	If one filled in the conformal boundary, with metric conformal
to (\ref{2}), with a metric analogous to (\ref{1}) but having a coordinate
different from $\tau = x^p$, say $x^k$ instead, having a nut at $r=r_0$,
then one would get a supergravity solution with
\be
\epsilon = \epsilon_k = -C_p \left({L\over L_k}\right)^{p+1}.
\label{7}
\ee
For $L_k \neq L_p$, this would correspond to a different state of the
gauge theory.

	Thus we see that if all of the $L_i$'s are different,
we get $p$ different AdS solitons that can fill in the conformal boundary
with representative metric (\ref{2}), one for each choice of the spatial
coordinate $x^k$ that is chosen to have the nut in the interior.
Each of these supergravity configurations has a different
scale-invariant energy given by Eq. (\ref{7}).

	If we chose an AdS soliton corresponding to an $L_k$ that is not
the shortest circle, then the scale-invariant energy would not be the minimum
possible value for that conformal boundary.
This would be a (rather trivial) counterexample to the positive-energy
conjectures of Horowitz and Myers
\cite{HM},
assuming that one measured the energy relative to the base metric
given by their Eq. (4.1) and had one of their $x^i$'s (without the nut)
having a shorter period than their $\tau$ that does have the nut.
(Of course, this would not be a counterexample if it is implicitly assumed
that the $x^i$'s have infinite range, as Horowitz and Myers \cite{HM} seem to do
except when they assume a finite $V_{p-1}$ in order to get a finite $E$.)

	In any case, one could trivially rephrase the Horowitz-Myers
conjectures to include the assumption that the base metric
has the period of the $\tau$ coordinate shorter than the period
of any of the other spatial coordinates transverse to $r$.
If these slightly revised conjectures are true, as I shall assume here,
then the lowest scale-invariant energy is the $\epsilon_k$
given by Eq. (\ref{7})
with $L_k$ chosen to be the shortest $S^1$ in the boundary metric (\ref{2}).
This would then be the scale-invariant ground state energy of the gauge theory:
\be
\epsilon_0 = \min_k{\epsilon_k} = -C_p \left({L\over\min{L_k}}\right)^{p+1}.
\label{8}
\ee
If we divide this by L, we get that the ground state energy
of the gauge field in the flat spatial $p$-torus of edge lengths $L_i$ is
\be
E_0 = -C_p \, {L_1L_2\cdots L_{k-1}L_{k+1}\cdots L_{p-1}L_p\over L_k^p},
\label{8b}
\ee
where $L_k$ is the minimum of the $L_i$'s.

	From dividing this energy by the volume, and from differentiating
the energy with respect to each of the edge lengths and dividing by
the transverse area, one can easily get that the stress-energy tensor
has only the following nonzero components,
in the flat coordinate basis used in the metric (\ref{2}),
and with $i$ indicating a spatial index not equal to the special index $k$
that labels the shortest $L_k$ (no sum on the repeated indices):
\be
T_{00} = - {C_p \over L_k^{p+1}},
\label{9}
\ee
\be
T_{ii} = + {C_p \over L_k^{p+1}},
\label{10}
\ee
\be
T_{kk} = - {C_p \, p \over L_k^{p+1}}.
\label{11}
\ee

	Thus we see that the energy density is negative,
there is a positive pressure of that same magnitude in each
of the periodic directions except for the shortest one,
and there is a negative pressure (tension) of $p$ times
that magnitude in the direction of the shortest periodic direction.
As expected, the trace of the stress-energy tensor is zero.

	This dependence on the spatial periodicities
of the stress-energy tensor of the large-N gauge theory
vacuum state in the spatial $p$-torus
gives a vacuum phase transition whenever the periodicities are changed
so that the direction of the shortest periodicity is switched.
The energy density, $T_{00}$, is continuous but has a discontinuity
in its derivative with respect to the length that either was or becomes
the shortest.
However, the pressures in the two directions that correspond to
what was and what becomes the shortest periodicity have discontinuities,
suddenly interchanging with the interchange of shortest lengths.
The strong coupling apparently makes the gauge theory vacuum highly sensitive
to the periodicity in the two shortest directions when they become equal.

	This sudden change in the stress tensor of the strongly coupled
gauge field vacuum is not similar to the smooth change in
the Casimir stress tensor for a weakly coupled gauge field,
so it is another feature of the difference between
strong and weak coupling, besides the famous factor of 3/4 (for $p=3$)
\cite{GKP}.

	Of course, for large but finite $N$, there would be no real
discontinuity in the stress tensor, and no real phase transition
for this system that is effectively in a finite cavity
(with periodic boundary conditions for the bosons
and antiperiodic boundary conditions for the fermions).
However, for large $N$, the stress tensor would change rapidly
with the two shortest periods when they are crossed,
as we shall discuss later.

	When one goes from the vacuum state to the thermal state at
a finite temperature $T$ for the strongly coupled gauge theory,
this is equivalent to making the Euclidean time periodic with
period $\beta = 1/T$, so the Euclidean metric for the gauge field
would simply be the $p+1$ torus $T^{p+1}$ with edge lengths
$\beta$ and the $p$ $L_i$'s.
In this case a slight modification of the analysis above
would predict that there should be a thermal phase transition
when $\beta$ drops below $L_k$, the shortest other periodicity.
This is indeed what has been found
\cite{SSW}.

	For $T < 1/L_k$, \cite{SSW} find one has a confinement phase,
with the expectation value of the temporal Wilson loop operator
being zero (in the large-$N$ limit).
The expectation value of the spatial Wilson loops along the spatial $S^1$'s
would be zero for all but the shortest $S^1$, which would have
a nonzero expectation value for its Wilson loop.
By the analysis above, using the AdS/CFT correspondence with only
the purely classical supergravity solutions,
one finds that the gauge field stress-energy tensor
has the form given by Eqs. (\ref{9})-(\ref{11}),
which thus does not change with temperature so long as it is below
the transition temperature $1/L_k$.
(This of course ignores the correspondence to the small effect of
thermal field fluctuations about the classical supergravity solution.)
Thus the strongly coupled gauge field is effectively frozen in its
confined ground state, with very low specific heat
(which would be slightly nonzero from the correspondence
with the thermal field fluctuations about the supergravity solution).

	For $T > 1/L_k$, \cite{SSW} find one has a deconfinement phase,
with the expectation value of the temporal Wilson loop operator
being nonzero.  Then the expectation value of all of the Wilson loops
over the spatial $S^1$'s would be zero (in the large-$N$ limit).
By a trivial extension of the analysis above,
one finds that the gauge field stress-energy has the form
\be
T_{00} = + C_p \, p \, T^{p+1},
\label{12}
\ee
\be
T_{ii} = T_{kk} = + C_p \, T^{p+1}.
\label{13}
\ee
(Here $T^{p+1}$ denotes the temperature raised to the power that is the
dimension of the spacetime in which the gauge theory is defined,
not a $p+1$ torus as has been my previous use of this expression.)

	This is exactly the same as the large-volume limit of a thermal gas
with 3/4 (for $p=3$) the number of degrees of freedom
as the weak coupling limit of the large-$N$ gauge field
\cite{GKP}.
However, I emphasize that this factor of 3/4 really applies only
when $\beta \ll L_k$.
When $\beta$ is comparable to $L_k$, the true thermal-Casimir stress-energy
tensor of the weakly coupled gauge field would be expected to change
slowly with the ratios of the periodicities, and not suddenly
as its components do in Eqs. (\ref{9})-(\ref{13}) for large $N$.
Thus it is not simply the factor of 3/4 that differs between the weak
and strong coupling limits, but also the more detailed dependence
on the periodicities.

	It may be of interest to give an improved approximation
for the stress-energy tensor for large but finite $N$
when the shortest periodicities are nearly equal,
which I shall do by using the zero-loop approximation
for the partition function for the supergravity theory
that is dual to the gauge theory.

	To shorten the expressions, I shall use $n \equiv p+1$,
the dimension of the spacetime in which the conformal gauge theory
lives, which in the thermal case (with periodic Euclidean time)
is the flat Euclidean $n$-torus with orthogonal periods and with
periodicity lengths $L_{\alpha}$ for $\alpha = 0,\ldots,n-1$,
with $L_0 = \beta$ and with the $n-1$ other $L_i$'s being as before.
For brevity, also define the $n$-dimensional volume
of the Euclidean spacetime to be
\be
V_n \equiv V_{p+1} = \beta V_p = L_0L_1\cdots L_{p-1}L_p,
\label{14}
\ee
and use
\be
C \equiv C_p \equiv C_{n-1}
  \equiv \left({4\pi \over n}\right)^{n} {\ell^{n-1}\over 16 \pi G_{n+1}}
  = {1 \over 4nG_{n+1}} \left({8\pi^2 (n-1)\over -n\Lambda}\right)^{n-1\over 2}.
\label{15}
\ee
From \cite{HM} one can see that for $n = p+1 = 4$, $C = (\pi^2/8)N^2$,
and from some examples of \cite{mal} for $n=3$ and $n=6$,
I would conjecture that for general $n$, $C \propto N^{n/2}$.

	Now, as discussed above, there are $n$ classical Euclidean supergravity
solutions with this Euclidean $n$-torus as their conformal boundary,
one for each choice of one of the $n$ $S^1$'s to be given a nut inside,
at which the periodicity length shrinks to zero to form,
along with the radial coordinate $r$, the center of a two-dimensional disk.
If it is the coordinate $\gamma$ that has the nut, then the action of that
solution is
\be
I_{\gamma} = - C {V_n \over L^n_{\gamma}}.
\label{16}
\ee
Then in the zero-loop approximation, this classical solution
makes a contribution to the partition function of
\be
Z_{\gamma} = e^{- I_{\gamma}}
             = \exp{\left({C V_n \over L^n_{\gamma}}\right)}.
\label{17}
\ee
Assuming the Horowitz-Myers conjectures
\cite{HM}
in the form revised above, which imply that these supergravity configurations
dominate the path integral, one has that
the total zero-loop partition function is
\be
Z = \sum_{\gamma=0}^{n-1} Z_{\gamma}
  = \sum_{\gamma=0}^{n-1} \exp{\left({C V_n \over L^n_{\gamma}}\right)}.
\label{18}
\ee
One can then say that each of the $n$ classical supergravity solutions
has probability
\be
P_{\gamma} = {Z_{\gamma} \over Z}
 = \exp{\left({C V_n \over L^n_{\gamma}}\right)}/
     \sum_{\delta=0}^{n-1}\exp{\left({C V_n \over L^n_{\delta}}\right)}.
\label{19}
\ee	

	Now using the toroidal symmetry of the metric
and differentiating the partition function by the nontrivial
parameters of the metric (the periodicity lengths $L_{\gamma}$)
gives the following stress-energy tensor (to zero-loop approximation,
which is good only for $C \gg 1$, and which ignores the correspondence
to the thermal field fluctuations in the dual supergravity theory
and other similar effects that would show up in a one-loop calculation
for that theory):
\be
T^{\alpha}_{\beta} = \sum_{\gamma=0}^{n-1}P_{\gamma}{C\over L^n_{\gamma}}
  \left(\delta^{\alpha}_{\beta}
  -n\delta^{\alpha}_{\gamma}\delta^{\gamma}_{\beta}\right),
\label{20}
\ee
where the Einstein summation convention is not used in the last term.

	Thus we see that for finite $N$, and hence for finite $C$
(which goes as a power of $N$, with the power apparently being half
the spacetime dimension $n$ in which the gauge theory lives),
there are no discontinuities in the stress-energy tensor
and no true phase transitions, which agrees with what one expects
on general grounds for a finite system.
However, for $C \gg 1$, the stress-energy tensor changes very rapidly
with the two shortest periodicities when they are very nearly equal.

	For example, when the inverse temperature,
$\beta \equiv 1/T \equiv L_0$,
is very nearly the same as the shortest spatial periodicity, say $L_k$,
and when all the other periodicities are significantly longer, then only
\be
P_0 \approx {1\over 2}\left[1+\tanh{
    \left({CV_n\over 2\beta^n}-{CV_n\over 2 L_k^n}\right)}\right]
\label{21}
\ee
and
\be
P_k \approx {1\over 2}\left[1-\tanh{
    \left({CV_n\over 2\beta^n}-{CV_n\over 2 L_k^n}\right)}\right]
\label{22}
\ee
are significantly different from zero,
and when they are both significantly different from zero,
they change very rapidly with $\beta$ and with $L_k$.
When one integrates $T_{00}$ over the spatial volume $V_{n-1}=V_n/\beta$,
one gets, for $\beta \approx L_k$,
\be
E \approx {C V_{n-1} \over L_k^n}(nP_0-1) = -E_0(nP_0-1),
\label{23}
\ee
where the negative $E_0$, given by Eq. (\ref{8b}), is the ground state energy
when the inverse temperature $\beta$ is taken to infinity.
Then one can calculate that the specific heat is
\be
{dE \over dT} \approx P_0 P_k \left({nCV_{n-1}\over L_k^{n-1}}\right)^2
              = P_0 P_k (-n L_k E_0)^2 = P_0 P_k n^2 C^2
	      \left({L\over\min{L_k}}\right)^{2n-2}.
\label{24}
\ee

	Since $C \gg 1$, and since $L$, the geometric mean of all of the $n-1$
spatial periodicities, is larger than $\min{L_k}$ (and can be much larger),
the specific heat can be very large when $P_0$ and $P_k$ are both comparable
to $1/2$.  Therefore, although there is not literally a phase transition
for finite $N$ (and hence finite $C$) and for finite $V_{n-1}/L_k^{n-1}$,
the stress-energy tensor can change very rapidly with the temperature
for large finite values of one or both of these quantities.

	Thus we can conclude that in the limit of infinite $N$,
a conformally invariant gauge theory
on a flat torus (with antiperiodic boundary conditions for the fermions),
dual to a supergravity theory in one higher dimension,
has vacuum and thermal states that are infinitely sensitive to the two shortest
periodicities of the torus when they are equal, giving a phase transition
when the two shortest lengths are interchanged.  This is analogous to what
was previously found
\cite{SSW} for the thermal phase transition when the inverse temperature crosses
the shortest spatial periodicity.  The phase transition involves a discontinuity
in the stress-energy tensor, in the components along the two shortest
periodicities (either both spatial, or one being the Euclidean time periodicity
for the thermal phase transition).

	I was introduced to the AdS soliton by Sumati Surya and Eric Woolgar
and had valuable discussions about it with them.
This work was supported in part by the Natural Sciences and Engineering
Research Council of Canada.



\baselineskip 4pt

\begin{thebibliography}{99}

\bibitem{HM} G.~T.~Horowitz and R.~C.~Myers,
Phys.\ Rev.\ {\bf D59}, 026005 (1999),
hep-th/9808079.

\bibitem{mal} J.~M.~Maldacena,  Adv.\ Theor.\ Math.\ Phys.\ {\bf 2}, 231 (1998),
hep-th/9711200.

\bibitem{gkp}S.~S.~Gubser, I.~R.~Klebanov, and A.~M.~Polyakov,
Phys.\ Lett.\ {\bf B428}, 105 (1998),
hep-th/9802109. 

\bibitem{W} E.~Witten,
Adv.\ Theor.\ Math.\ Phys.\  {\bf 2}, 253 (1998),
hep-th/9802150. 

\bibitem{SSW} S.~Surya, K.~Schleich, and D.~M.~Witt,
Phys.\ Rev.\ Lett.\ {\bf 86}, 5231 (2001),
hep-th/0101134.

\bibitem{GH} G.~W.~Gibbons and S.~W.~Hawking,
Commun.\ Math.\ Phys.\ {\bf 66}, 291 (1979).

\bibitem{GKP} S.~S.~Gubser, I.~R.~Klebanov, and A.~W.~Peet,
Phys.\ Rev.\ {\bf D54}, 3915 (1996), 
hep-th/9602135.


\end{thebibliography}
\end{document}